# Administration 4.0: Administrative informatics as a customized and necessary educational platform for a modern IT-supported federal administration

Uwe M. Borghoff [1], Nicol Matzner-Vogel [2] and Siegfried Rapp [3]

**Abstract:** Digitalization is conquering and stressing out the federal administration. Using selected large-scale ICT projects, we show how complex and interdisciplinary the tasks are. The federal administration's IT strategy requires well-trained specialists for all defined fields of action. This scarce resource is increasingly being trained academically in separate, tailor-made degree courses that are developed specifically for the needs of the German ministries and authorities. We use the example of administrative informatics courses to explain their necessity and success story. Using a Bachelor's/Master's program developed by the authors for the ITZBund and the BMF[4], we look at a concrete implementation and justify two of our design decisions in the development of the course, namely *transdisciplinarity* and *design thinking*. We adopt a German perspective throughout the paper[5]. However, the conclusions also apply to other countries.

**Keywords:** digitalization, large-scale ICT projects, IT strategy of the federal administration, central fields of action, recruitment of young talent, administrative informatics courses.

## 1 Introduction

Digitalization is advancing in all areas of work, not only in business and industry, but also in the public sector [LEW21]. Public administrations are dependent on the smooth use of modern information technology, which is why information technology is becoming increasingly important. However, a modern administration is not sustainable without qualified personnel. Due to the enormous demand from the private sector for IT workers of all kinds, federal and state authorities and ministries are facing the challenge of recruiting young talent in a targeted manner by meeting the demand for junior staff in specially designed degree courses with a specialist focus on administrative informatics. This is not least a response to the experience gained from large, cost-intensive ICT (information and communication technology) projects run by the federal government, thereby contributing to a sustainable digitalization initiative for the Federal Republic of Germany. Administrative IT specialists develop, control and maintain IT processes that map the real administrative processes as accurately as possible. For this reason, in addition to the pure IT content, management knowledge and basic knowledge of the relevant legal situation and common administrative processes must also be taken into account when developing a degree program. The first of our design decisions, *transdisciplinarity*, also plays a cross-sectional role here. We go into this in more detail in section 4.3.

---

[1] Universität der Bundeswehr München, Institute for Software Technology & Campus Advanced Studies Center (CASC), 85577 Neubiberg, uwe.borghoff@unibw.de, https://orcid.org/0000-0002-7688-2367.

[2] Universität der Bundeswehr München, Campus Advanced Studies Center (CASC), 85577 Neubiberg, nicol.matzner@unibw.de, https://orcid.org/0009-0008-9525-0008.

[3] Universität der Bundeswehr München, Central Administration, 85577 Neubiberg, siegfried.rapp@unibw.de, https://orcid.org/0009-0004-3296-0732.

[4] Federal Ministry of Finance.

[5] This paper was presented at the practitioner session of the RVI 2023 conference in Dresden, Germany, on 26-27 October, see also https://www.rvi23.de/pages/programm.html (accessed 09.12.2023).

In order to make the federal administration competitive in the face of globalization and digitalization, also in comparison with the private sector, the federal government has decided on a series of measures to advance the digitalization of the administration. The federal administration needs flexibly adaptable, future-oriented and secure IT technology. Until now, the federal government's information technology, and therefore that of the respective departments and agencies, has been largely decentralized. This meant that the information technology was equipped differently in terms of hardware, software and security architecture and that there were a large number of isolated solutions. Uniform and centrally controlled information technology offers great opportunities for a modern state. As digitalization increases, so does the need for secure digital and citizen-oriented services.

Citizen orientation in particular is putting the existing distributed administrative processes to the test. In the future, citizens should be offered administrative processes (centralized and, above all, integrated) that do not focus on local official responsibilities and their own internal processes (and often isolated database systems), but rather on the user perspective. The repetitive entry of data records, which would already exist in the (overall) administration system if it were federated and integrated, would no longer be necessary. User satisfaction and acceptance would increase. This is precisely the aim of the Online Access Act (OZG) and the Register Modernization Act (RegMoG) together with the associated measures, which we will discuss in more detail in sections 2.2 and 2.3. The second of our design decisions, *design thinking,* is based on focusing on the user perspective. We will also go into this in more detail in section 4.3.

## 2 Public sector and large-scale ICT projects

In this section, we look at some major ICT projects in the public sector. It will become clear that the skills taught in administrative informatics - computer science, (administrative) management, administrative theory and law - are urgently needed.

Specifically, we look at the federal IT consolidation, the OZG measures and the register modernization as well as the Herkules project.

### 2.1 Federal IT consolidation

In 2015, the federal government adopts a landmark decision. It concerns the federal IT consolidation project, which aims to centralize and standardize most of the information technology. The aim is to achieve efficient, economical, stable and future-proof operations. IT procurement has also been bundled in order to become more efficient. The federal administration's expenditure on IT investments as well as IT contracts and IT services was estimated at around €5.4 billion in 2021 (2016: €1.8 billion) in the Federal Budget Act[6]. In 2020, the federal administration had 68,000 server instances, 345,000 end devices and 14,000 specialist applications. The number of employees in information technology doubled between 2016 and 2021 to around 8,000[7].

This gigantic IT landscape requires trained and well-educated personnel at all levels. Ogonek *et al* [Og16] have defined 19 roles and associated competencies for this, which are comprehensively addressed in most of the study programs mentioned in section 4.

### 2.2 OZG measures

The Online Access Act[8] (OZG) obliges the federal and state governments to offer their administrative services electronically via administrative portals by the end of 2022 at the latest. To this end, more than 6,000 administrative services will be identified and summarized in 575 so-called OZG service bundles. The OZG implementation catalog[9] contains 14 subject areas, which in turn cover 35 life situations and 17 business situations for companies. The *federal* digitalization programme includes all services that are the sole responsibility of the federal government in Berlin, Germany. The Federal Ministry of the Interior and Community has been assigned the lead responsibility. In addition, there is the Federal Digitization Programme for the services to be provided with regulatory competence for the *federal states* (Bundesländer) and local authorities.

Topics include law and order, the environment, taxes and customs, immigration and emigration, health, family and children, and business management and development. These include services ranging from birth certificates, health services, the issuing of ID cards and building permits to the central trade register. The federal government is responsible for and regulates 115 service bundles. In October 2020, an OZG dashboard was made available online, providing an overview of which administrative services are currently available in Germany.

---

[6] Act on the Adoption of the Federal Budget for the 2021 Financial Year (Budget Act 2021), December 21, 2020 (includes all fixed IT titles and title groups from all individual plans as well as other IT-related titles).

[7] Bundesrechnungshof - Report to the Federal Chancellery and the Federal Government Commissioner for Information Technology pursuant to Section 88 (2) BHO; Federal IT management; GZ: VII 3 - 2020 - 0088 dated November 24, 2021.

[8] Act to improve online access to administrative services (OZG) of August 14, 2017.

[9] IT Planning Council, OZG implementation catalog, Digital administrative services within the meaning of the Online Access Act, 1st edition, version 0.98; Berlin, April 2018.

To implement the OZG, a so-called maturity model[10] was developed, which defines the degree of digitization of an administrative service in order to meet the legal requirements of the OZG. The online availability of an administrative service is mapped on a scale from 0 to 4. 0 stands for no digitization, 1 for online availability of the service description, 2 for online application possible in principle, but without transmission of evidence and 3 for digital online processing including all evidence and delivery of the notification in digital form. With 4, the so-called *once-only principle is* introduced online. For application processes and context integration, this means that applications and objection procedures are possible online. The user's master data is taken from any user or company account in the portal network. We will go into this in more detail in section 2.3.

Achieving a high level of maturity requires not only technical expertise, but also in-depth knowledge of administrative law and the entire administrative process. The mapping of various administrative acts, even among several parties and authorities, shows the full complexity. Administrative informatics courses should therefore focus on administrative procedures and the various administrative acts, including public law contracts, particularly in administrative law.

A review of the OZG revealed that only a small proportion of administrative services had been digitized by the end of October 2022. Only 33 of the 575 service bundles with the underlying administrative services were available across the board[11]. The Bundesrechnungshof[12] specified this for the federal government in its audit reports to the effect that only 10% of the digitizable services were available in accordance with the OZG. The Federal Minister of the Interior and Community assumes in the information provided by the Federal Government[13] to Parliament on May 26, 2023 that 97 out of 115 OZG services are offered digitally; however, it is pointed out that a large number of these are not yet offered across the board. A planned amendment to the OZG is expected to bring significant improvements.

### 2.3 Register modernization

Register modernization (or the Register Modernization Act) attempts to bring a tried and tested IT principle from the database world into the world of administration: *once-only* (maturity level 4). With the help of an overarching classification feature - a unique identification number (ID no.) - data and evidence are to be transmitted digitally and, above all, across administrations without the need to repeatedly re-enter or resubmit data already recorded in the various administrative databases. The obvious use of a suitable ID number from a database perspective, namely the tax identification number (tax ID), seemed to be found quickly. The tax ID is a random 11-digit number that is (ideally) uniquely assigned to a person. When it is randomly generated, a check digit is also calculated and other properties (permissible digit multiplicities etc.) are taken into account, which make it highly likely that simple transposed digits and typing errors will be detected. Even though it was initially developed purely for tax purposes, at first glance it fulfills the requirements of an ID number within the meaning of the RegMoG. Unlike, for example, the personal identification number of the German Bundeswehr (PK number), which contains the date of birth, the first letter of the surname and information about the district military office or the registration district, the tax ID is not "talkative" and does not allow any conclusions to be drawn about the person.

---

[10] Federal Ministry of the Interior and Community, OZG Federal Digitization Programme - Maturity Model, Version 1.1, page 5.
[11] National Regulatory Control Council, Annual Report 2022, Bureaucracy reduction in the new era - relieving the burden on citizens, business and administration now.
[12] Bundesrechnungshof, Report to the Budget Committee of the German Bundestag, Implementation of the Online Access Act, Control and Coordination, Gz: VII 5 - 0001755 of March 29, 2023, page 15.
[13] German Bundestag, 20th legislative period, briefing by the Federal Government, key points for a modern and future-oriented administration; printed matter 20/7115 of May 26, 2023.

Register modernization is an excellently suited subject for any administrative informatics course, as it requires transdisciplinary thinking. On the surface, this is a database topic that is easy to explain, as described above. However, by summer 2020 at the latest, it was clear that the Conference of Independent Federal and State Data Protection Supervisory Authorities [Re20] and the German Informatics Society (GI) had serious concerns. The data protection experts pointed out the risk that personality profiles could be created through a register comparison, i.e. through the potentially abusive combination of content from different databases. The federal government must counter this risk with the greatest possible transparency and traceability of (end-to-end *encrypted*) cross-administrative data exchange (keyword: *data cockpit*). The data protection experts also believe that the risk of misuse will increase in the future, as this simple (global) ID number will also be widely used in business. They are therefore calling for sector-specific personal identification numbers, which in turn uniquely identify a person. A one-sided (state) comparison should at least be made more difficult[14]. The GI rejects a global ID number as unnecessary and instead recommends the development of a comprehensive, state-of-the-art identity management architecture for public administration [GI20].

### 2.4 Herkules

There are probably few major federal ICT projects that have been reported on as critically as the Herkules project. Between 2006 and 2016, the Bundeswehr modernized its entire non-military ICT equipment and also standardized the interfaces and software platforms. The most important software decisions concerned the introduction of SASPF (SAP's Standard Application Software Product Families) and the nationwide installation of a public key infrastructure for encrypting and digitally signing electronic documents.

---

[14] However, this cannot be completely prevented. Andreoli *et al*. presented a theoretical model 30 years ago [ABP94], which makes it possible to combine the contents of heterogeneous (freely accessible) databases, even if no global or domain-specific IDs are known in today's linguistic usage. Based on this, Borghoff and Schlichter [BS96] co-developed a prototype which was later marketed by Xerox under the name *AskOnce* (sic!) and taken over by Documentum in 2004.

All three phases of the project (actual operation (migration phase), mixed operation (integration phase) and target operation) were scientifically examined with regard to customer satisfaction. Krampe [Kr12] from the Bundeswehr Institute of Social Sciences writes in his report published in 2012 that the development of customer satisfaction after completion of the integration phase is viewed ambivalently (see also [Kr11]). On the one hand, satisfaction with BWI customer services increased further, but satisfaction with IT equipment and network operation fell.

There was a political discussion about the exploding costs and, not least, about the continuation of the project in target operation. The latter was resolved by merging BWI Informationstechnik and BWI Systeme to form BWI, based in Meckenheim, Germany, a wholly owned federal company with limited liability[15]. BWI has been developing and operating the Bundeswehr's ICT infrastructure for an unlimited period since 2017. It also provides these services to other federal ministries.

Despite all criticism, we see the Herkules project overall as a milestone in increasing ICT efficiency and in the positive change in the Bundeswehr's digitization philosophy. The concept proposals made very early on by Borghoff *et al* [BHH04] for the digitization of documents, especially service regulations, went a long way. With Herkules

- standardization in IT equipment and the associated service,
- extended process automation and
- further centralization of admin processes can be achieved (see also [Ri12]).

This is one of the reasons why the transfer of experience gained and the Herkules functional scope are important topics in the training of junior IT staff. They also influenced the decisions in the development of the federal administration's IT strategy, which we will now discuss in more detail.

## 3  IT strategy of the Federal Administration and key areas of action

The Federal Administration's IT strategy for 2017-2021 was promulgated by the IT Council[16] and subsequently published by the Federal Government Commissioner for Information Technology. It was published as the federal IT strategy for 2017-2021 and is constantly updated[17] (most recent version 2022). The IT strategy defines overarching goals and fields of action for the federal administration's information technology.

---

[15] https://www.bwi.de (accessed May 16, 2023).
[16] IT Council, resolution no. 2017/5 of January 19, 2017.
[17] https://www.cio.bund.de/Webs/CIO/DE/digitaler-wandel/it-strategie/it-strategie-node.html (accessed August 24, 2023).

Ten strategic goals are documented for the further development of information technology in the Federal Administration.

1. *Effectiveness and quality*: Automated IT-supported processes require a very high degree of professionalism. This means that the federal administration's IT systems must be procured, operated and further developed in a customized manner. A modular structure and standardization must be prioritized and brought together in a few data centers.

2. *Digital administration*: The digitalization of society is progressing steadily in all areas and also affects the federal administration. Administration, business and society must be networked with each other. Administrative processes must be made available digitally, regardless of time and place. Orders, notifications and invoicing should be processed electronically in future.

3. *Future viability and openness to innovation*: Innovative IT solutions from the federal administration have a competitive impact on Germany as a business location. Knowledge exchange and communication should be supported by central platforms. This requires openness towards new technologies.

4. *Information security and data protection*: The federal administration must protect itself against hacker attacks and cybercrime. Measures to protect information and personal data are therefore absolutely essential. Digital administrative procedures, electronic files and processes must be protected against manipulation and identity misuse in particular.

5. *Attractiveness as an employer*: The demand for IT specialists in the Federal Administration has increased enormously. Career models in the various career paths must be presented. At the same time, family-friendly working hours, factors to improve the work-life balance, further training opportunities from a lifelong learning perspective and an attractive employer environment are expected.

6. *Economy and cost efficiency*: Economy and cost efficiency are two key pillars for the implementation of the federal government's IT strategy. Efficiency and effectiveness are to be increased by exploiting synergies, taking into account the principle of cost-effectiveness.

7. *Inclusion and accessibility*: IT systems can provide support, but they can also be a barrier. All digital processes and services must be accessible. The same applies to employees' workplaces.

8. *Environmental compatibility and sustainability*: Sustainability aspects must also be taken into account in all areas with regard to climate change. The sustainability strategy must be observed[18].

---

[18] Federal Government, German Sustainability Strategy, new edition 2016, further development 2021.

9. *Cooperation*: Federal IT must consider various levels of administration and the private sector. Possible collaborations include various administrative levels in the EU, the federal government and the federal states as well as companies and numerous other national and international institutions.

10. *Controllability and manageability*: In order to ensure that IT services are provided in line with requirements at all times, a higher-level management system, including IT controlling, is required. These tasks are performed across the board by the ITZ-Bund.

Key areas of action are defined in order to implement these strategic goals. The fields of action are constantly updated and updated at least every five years in order to keep pace with political and technological developments and requirements. Administrative informatics courses naturally address the central field of action of *digital skills* (previously: *development of IT personnel*). However, they are also required to include the areas of *cloud computing*, *IT security* and *digital infrastructures* (such as computer networks) in their curriculum. In our opinion, the field of *e-government* (including digital administration) must be integrated into the degree programs. The digitalization of administrative work is not possible without a sound knowledge of administrative processes, administrative procedures and administrative law. Only in this way is it possible, for example, to process digital notifications and objection notices completely online in order to meet the requirements of the Online Access Act. At the same time, the legal provisions on IT security and data protection must be complied with in an application-oriented manner. Cost-effective procurement in terms of economic efficiency[19] and public procurement law should also be comprehensively communicated.

We therefore strongly recommend that module handbooks and examination regulations are reviewed regularly, at least every five years, with a view to the current fields of action of the Federal IT Strategy and updated if necessary.

# 4 Selected administrative informatics courses for public authorities and ministries

The enormous demand from public authorities and ministries, particularly for junior IT staff, is increasingly being met by specially designed degree courses with a focus on administrative informatics.

---

[19] Cf. section 7 of the Federal Budget Code (BHO).

| Institution and location | Procedure | Special features |
|---|---|---|
| HS Bund - Brühl & Münster [Ho23a] | Diploma (FH), 2 years + 3 practical phases, no ECTS | Dual study program, 24 months at the Federal University of Applied Sciences for Public Administration, 12 months practical training with the employer, training contract required |
| Duale HS Gera-Eisenach [Du23] | B. Eng./ B. Sc., 3 years, 180 ECTS | dual, training contract with practice partner required |
| HS Hannover [Ho23c] | B. Sc., 7 semesters, 210 ECTS | Full-time, eligible for a scholarship, 4-week pre-study internship in a relevant authority |
| HS Hof [Ho23e] | Diploma (FH), 6 semesters, no ECTS | Propaedeutic course in mathematics, 6th semester at the employer and the University of Applied Sciences for the Civil Service in Bavaria, Germany |
| University of Applied Sciences for Public Administration Rhineland-Palatinate - Mayen [Ho23b] | B. A., 3 years, 180 ECTS | dual study program, 21 months at the university, 15 months practical training with the employer, Training contract required |
| UniBw Munich [Un23] | B. Sc., 3 years with two 4-week internship phases with the employer, 180 ECTS | Full-time in trimesters; followed by a part-time university Master's degree as a career advancement course, 3 years à 40 ECTS. Training contract required. Exclusively for ITZBund and BMF |
| HS Harz - Wernigerode [Ho23d] | B. A., 7 semesters, at least 180 ECTS | dual, Harz University of Applied Sciences, focus on ERP systems (*Enterprise Resource Planning*, e. g. SAP), Bachelor's thesis in the 7th semester with a practice partner. Study or scholarship contract required |

Table 1: Selected administrative informatics courses for public authorities and ministries

Table 1 provides a brief overview of some relevant administrative informatics degree programs in Germany. In addition to purely dual courses offered by public authorities, there are also courses that can be studied full-time. Further offers are listed on StudyCheck[20]. We deliberately only look at administrative informatics degree programs that were designed (more or less) exclusively for state and federal authorities or ministries. We do not include undergraduate and consecutive Business Informatics or pure Informatics degree programs, which can be studied at almost all German universities. There are three reasons for this: Firstly, the authorities and ministries require the skills taught in the areas of computer science, (administrative) management, administrative theory and law, as we have already described. Secondly, all of the administrative informatics degree programs considered (largely) dispense with mathematics and theoretical computer science components, which increases the potential field of applicants. Finally, there is a well-founded fear that the attraction of the economy to business and IT specialists and the associated salary pressure could lead to an early and excessive exodus from the public sector.

We will now take a closer look at our degree program [Un23]. This is offered both as a full-time Bachelor's degree course and as a part-time Master's degree course based on it. The ordinance on promotion to the higher technical or non-technical federal administrative service for this Master's degree course was published in the Federal Law Gazette on March 24, 2023.

### 4.1 The Bachelor's degree program in Administrative Informatics (B. Sc.) at the University of the Bundeswehr Munich

The aim of the course is to impart information technology, business, management and legal knowledge explicitly tailored to the special features of public administration. It is aimed at candidates for the higher technical administrative service in the area of responsibility of the Federal Ministry of Finance. The recruiting authority is the ITZBund. In particular, the course is intended to teach the ability to independently analyze specific application areas and requirements for information systems and to design, implement and operate solutions according to the current state of technology and science. This combines a sound education in the field of ICT (information and communication technology) with administrative science.

The course comprises three pillars:

- Computer science
- Administrative management
- Administrative theory and law

The focus of the course is on computer science. In the core computer science subjects, students deal with the fundamentals of computer science, software development and programming. Aspects of data protection and IT security, IT law and the classification of individual issues in the overarching context of digital transformation round off the module canon in the first pillar.

---

[20] https://www.studycheck.de/studium/verwaltungsinformatik (accessed May 16, 2023).

In addition to information technology, economic, management-oriented and legal principles of administration are also taught. In the second pillar, students are familiarized with the basics of project management, quality management and the efficient design of processes. The teaching of scientific methods and basic economic skills and their application in internal accounting and controlling enable students to reflect on their professional activities from a business management perspective. The Public Management module forms the framework for the classification of various issues in the overarching process of modernizing administrations in the context of digitalization.

The third pillar enables students to know and apply legal principles. Administrative law and public service law are at the heart of this part of the course, but constitutional, European and civil law as well as budgetary law with its application to economic efficiency studies also play a key role in determining the legal framework for the professional activities of administrative IT specialists.

The course is a successful model and is enjoying increasingly high student numbers. The pilot course had 23 students in 2020. This number grew to 35 in 2021, 63 in 2022 and is expected to reach 70 in the following years.

### 4.2 The Master's degree program in Administrative Informatics (M. Sc.) at the University of the Bundeswehr Munich

In contrast to the Bachelor's degree, the Master's degree is designed to be taken alongside work in order to ensure a good work-study balance and to achieve a close link between the academic skills gained and their application directly in the sending authorities. It takes into account the requirements of the Federal Administration's IT strategy and includes modules in law, business administration, business informatics and, of course, IT. The computer science components are essentially practical, technical and applied computer science. Once again, theoretical computer science plays no role in the Master's degree course.

### 4.3 Transdisciplinarity and design thinking in the degree program

As we have just seen, different scientific disciplines are taught in both the Bachelor's degree and the part-time Master's degree.

During the development of the course, the team of lecturers focused on the so-called *transdisciplinary* approach. This means that the individual disciplines should not stand next to each other in isolation, but should influence each other, enable cross-referential thinking and promote the necessary transfer from one discipline to another. This is particularly important in programming and, on a larger scale, in software engineering. Albrecht *et al* [ANB22] can show that transdisciplinary software engineering can lead to a common working language throughout the entire software development process and to a better understanding of customer expectations as early as the requirements analysis stage. After all, customers are not usually IT specialists, but live in their own discipline. Such ideas flow directly into the regular lecturer conferences. Often an impulse also comes from the customer themselves.

For example, the ITZBund, as a demand-driven institution, placed value on *design thinking* [Br08] for all students at a very early stage in the development of its degree programs. This can also be seen as a further approach to transdisciplinarity. Customer interests are integrated into the (IT) development process at a very early stage and in a structured manner. The Institute for Software Technology[21] has been offering a so-called programming project for various degree courses for many years. For most students, this is their first contact with a larger software development project. Student teams (usually 6-8 people) go through the phases of requirements analysis, specification creation, design, development, testing and delivery (maintenance and operation are omitted) in different roles along the waterfall model. The waterfall model was deliberately chosen as it is probably the simplest and most accessible process model. Complicated models such as the V-model or agile methods (*Scrum* etc.) are explained in theory but not practiced in teams at this early stage. We are also trying to apply the very good experiences with the computer science programming project - even during the pandemic [BMM23] - to administrative informatics. When selecting the modules, a great deal of consultation was necessary with the client and the team of lecturers, as we are not dealing purely with computer scientists. The selection decisions made are repeatedly put to the test in annual review meetings and are subject to the university's usual evaluation procedures.

## 5   Summary and outlook

### 5.1   Tailor-made educational platform

First of all, a word about why we chose the term *educational platform* in the title. Our degree program [Un23] is the substructure of an integrated educational platform in which an undergraduate Bachelor's degree program, a part-time Master's degree program and practical training fully integrated into the degree programs interact in the form of internship phases lasting several weeks and monitored by the university and the associated evaluation processes[22][23]. Through our alumni network, we are also trying to use this platform for lifelong learning opportunities in the future (see also Goal 5 of the Federal Administration's IT strategy from Section 3).

---

[21] https://www.unibw.de/inf2 (accessed May 16, 2023).
[22] https://publicwiki.unibw.de/display/DAT/Studium (accessed August 16, 2023).
[23] https://www.unibw.de/casc/praktikumsleitfaden_vit.pdf/download (accessed August 20, 2023).

The catalog of possible internship fields and internship placements negotiated with the providers together with the binding specifications of the internship officer of the University of the Bundeswehr Munich with the instructors in the authority on internship activities suitable for higher education and their examination-relevant documentation was noted with approval by the accreditation agency and the entire degree program was accredited without conditions.

### 5.2 Further development and outlook

Ministerially or officially supported - or, to be more precise, co-designed - degree programmes such as administrative informatics are successful models for alleviating the shortage of IT specialists.

All of the degree programmes considered are comparable in terms of content and support Germany's sustainable digitalization initiative. However, the degree programs are also well-behaved, conservative and "German". We currently see three focal points in curricular development.

On the one hand, there is often a lack of understanding of the traceability of administrative actions, even in the presence of electronic artifacts such as e-files; see also Schilling [Sc16].

On the other hand, disruption through artificial intelligence is only dealt with superficially, if at all. Lucke [Lu19] and Nay [Na22] provide important impulses in this regard. There is potential for development here, which should also be used to sharpen the content and profile of degree programs. The central field of action of the IT strategy Bund *Technological Change,* including artificial intelligence in administration, urgently needs to be integrated with a view to the currently booming generative AI approaches.

The digitalization initiative does not end at the German border. Thirdly, we are trying to focus even more on international projects in our Master's program in the future. Register modernization, for example, also has (at least) a European dimension.